\documentclass[aps,prL,superscriptaddress]{revtex4}
\usepackage{amsmath}
\usepackage{amssymb}
\usepackage{bm}
\usepackage{epsfig}
\usepackage{graphicx}
\usepackage{color}
\usepackage{dcolumn}
\usepackage{bm}
\begin{document}
\title{Resonant bending of silicon nanowires by light}
\author{Evgeny Bulgakov}
\author{Almas F. Sadreev}
\affiliation{Kirensky Institute of Physics, Federal Research
Center KSC SB RAS, 660036 Krasnoyarsk, Russia}
\date{\today}
\begin{abstract}
Coupling of two dielectric wires with rectangular cross-section
gives rise to bonding and anti-bonding resonances. The latter is
featured by extremal narrowing of the resonant width for variation
of the aspect ratio of the cross-section and distance between
wires when the morphology of the anti-bonding resonant mode
approaches to the morphology of the Mie resonant mode of effective
circular wire with high azimuthal index. Then plane wave resonant
to this anti-bonding resonance gives rise to unprecedent
enhancement of the optical forces up to several nano Newtons per
micron length of wires. The forces oscillate with angle of
incidence of plane wave but always try to repel the wires. If the
wires are fixed at the ends the optical forces result in elastic
deflection of wires of order $100 nm$ for wires's length $50\mu m$
and the light power $1.5mW/\mu m^2$.
\end{abstract}
\maketitle


\section{Introduction}
The response of a microscopic dielectric object to a light field
can profoundly affect its motion. A classical example of this
influence is an optical trap, which can hold a particle in a
tightly focused light beam \cite{Ashkin1986}. When two or more
particles are present, the multiple scattering between the objects
can, under certain conditions, lead to optically bound states.
This is often referred to peculiar manifestation of optical forces
as optical binding (OB), and   it   was first   observed by Burns
et   al.  on   a   system   of two  plastic   spheres in water  in
1989 \cite{Burns1989} and after by other scholars
\cite{Chaumet2001,Tatarkova2002,Mohanty2004}. Optical binding
belongs to an interesting type of mechanical light-matter
interaction between particles at micro-scale mediated by the light
scattered by illuminated particles. Depending on the particle
separation, OB leads to attractive or repulsive forces between the
particles and, thus, contributes to the  formation  of stable
configurations of particles. The phenomenon of OB can be realized,
for example, in dual counter propagating beam configurations
\cite{Tatarkova2002,Gomez-Medina2004,Metzger2006,Dholakia2010}.

It is clear that excitation of  the resonant modes with high $Q$
factor in dielectric structures by light results in large
enhancement of near electromagnetic (EM) fields and respectively
in extremely large EM forces proportional to squared EM fields.
First, sharp features in the force spectrum, causing mutual
attraction or repulsion between successive photonic crystal layers
of dielectric spheres under illumination of plane wave has been
considered by Antonoyiannakis and Pendry
\cite{Antonoyiannakis1997}. It was shown that the normal force
acting on each layer as well as the total force acting on both
layers including the optical binding force follow these
resonances. It was revealed that the lower frequency bonding
resonance forces push the two layers together and the higher
frequency anti-bonding resonance pull them apart. Later these
disclosures were reported for coupled photonic crystal slabs
\cite{Liu09} and two planar dielectric photonic metamaterials
\cite{Zhang2014} due to existence of resonant states with infinite
$Q$ factor (bound states in the continuum (BICs)). Recently
Hurtado {\it et al} \cite{Hurtado2020} have shown that excitation
of quasi BICs in a dimerized high-contrast grating with a
compliant bilayer structure stimulates considerable forces capable
for structural deformations of the dimer.

\begin{figure}
\includegraphics*[width=8cm,clip=]{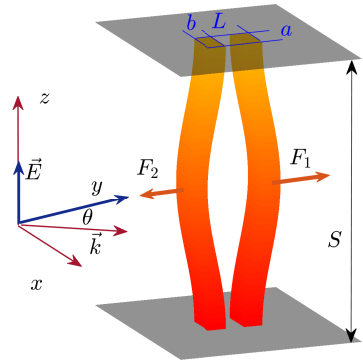}
\caption{Two long dielectric wires fixed between two parallel
planes are bending under illumination of plane wave incident by
angle $\theta$ in the plane $x$ and $y$. Arrow shows the electric
vector of the incident wave with power $P_0=1.5mW/\mu m^2$.
$a=1.19\mu m, b=0.592\mu m, S=50\mu m$.} \label{fig1}
\end{figure}

However it is remarkable, even two particles can demonstrate
extremely high $Q$ resonant modes owing to avoided crossings. The
vivid example is avoided crossing of whispering-gallery modes
(WGM) in coupled microresonators  which results in extremely high
$Q$ factor \cite{Povinelli2005,Benyoucef2011}. As a result an
enhancement of the OB force around of hundreds of nano Newtons
between coupled WGM spherical resonators takes place in applied
power $1mW$ \cite{Povinelli2005}.  After a few years Wiederhecker
{\it et al} \cite{Wiederhecker2009} demonstrated a static
mechanical deformation of up to 20 nm in double silicon nitride
rings of 30 $\mu m$ diameter by illumination of milliwatts optical
power. In 2005 Povinelli {\it et al} \cite{Povinelli2005a}
calculated forces between two parallel, silicon wires of square
cross-section as shown in Fig. \ref{fig1} (a) caused by
electromagnetic (EM) waves propagating along the wires with
frequencies below the light line. Both attractive and repulsive
forces, determined by the choice of relative input phase, the
forces were found large enough to cause displacements 20 nm of
wire with length 30$\mu m$. However these optical binding (OB)
forces are caused by evanescent EM fields which are exponentially
weak between the wires that requires a considerable input power
1W/$\mu m^2$ \cite{Povinelli2005a}. In the present letter we
consider scattering of plane wave by parallel wires in the
resonant regime. Scheme of illumination is shown in Fig.
\ref{fig1} (a). That demands  incident power of only 1.5 mW/$\mu
m^2$ in order to result considerable deflection  of wires of order
100 nm for the wires of length $50 \mu m$.

We show that for two-parametric variation (the aspect ratio and
distance between wires) the system of two wires acquires
anti-bonding resonant modes with extremely high $Q$ factor. That
happens when a morphology of the anti-bonding resonant mode
becomes close to the morphology of the Mie resonant modes with
high azimuthal index of effective cylindrical wire. If the plane
wave with power $1.5 mW/\mu m^2$ is capable to excite such an
anti-bonding resonant mode optical forces reach a value till one
nano Newton per micron length of wires. As a result wires of
enough length are bending as shown in Fig. \ref{fig1}.
\section{Mie resonance in the system of two parallel nanowires}
\begin{figure}[ht!]
\includegraphics[width=9cm,clip=]{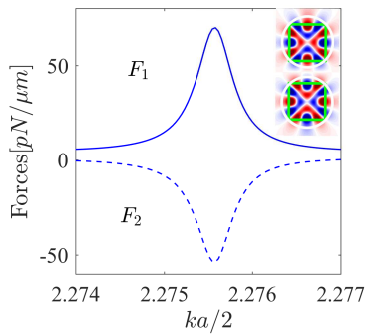}
\caption{The OB forces acting on wires of squared cross-section
with $a=b=1\mu m$ versus frequency of plane wave incident normally
$\theta=0$ with power $1.5mW/\mu m^2$.} \label{fig2}
\end{figure}

The behavior of resonances of resonances of isolated nanowire as
dependent on aspect ratio of the cross-section $a/b$ was studied
by Huang {\it et al} \cite{Huang2019} in aim to optimize the $Q$
factor of the nanowire by use the same strategy of avoided
crossing as it was used in papers \cite{Wiersig2006,Rybin2017}.
Therefore it is reasonable to start a consideration of optical
binding (OB) force of two wires each of them is optimized by the
$Q$ factor which reaches maximal amount 2730 for silicon wire with
the refractive index 3.48 at $a=b$. The corresponding resonant
mode of each wire shown in inset of Fig. \ref{fig2} has a
morphology of the Mie resonant mode with rather high azimuthal
index $m=6$ (close to whispering gallery mode) that explains so
small radiation losses. The corresponding optical forces, acting
on each wire are shown in Fig. \ref{fig2}. One can see that the OB
force tends to repel the wires and reaches a magnitude 70 pico
Newtons per micron.

\begin{figure}[ht!]
\includegraphics[width=7cm,clip=]{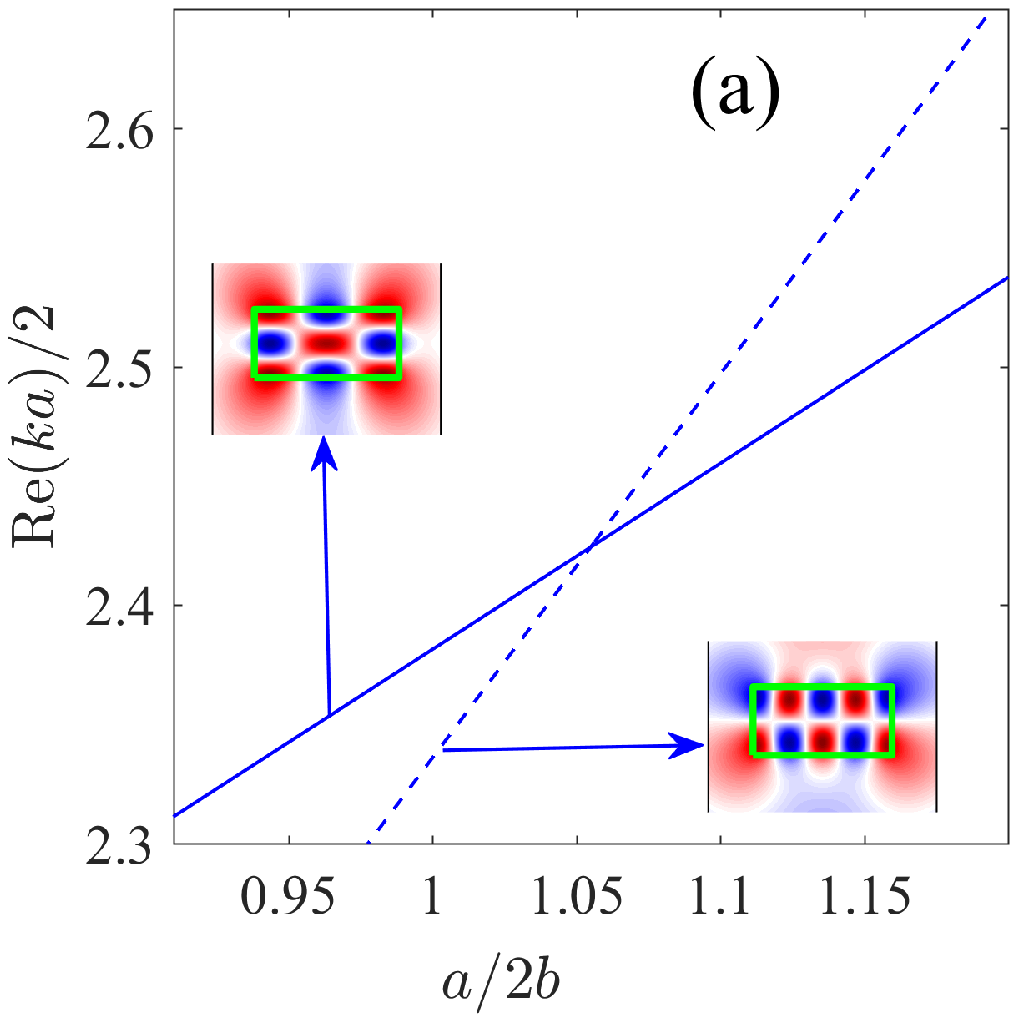}
\includegraphics[width=7cm,clip=]{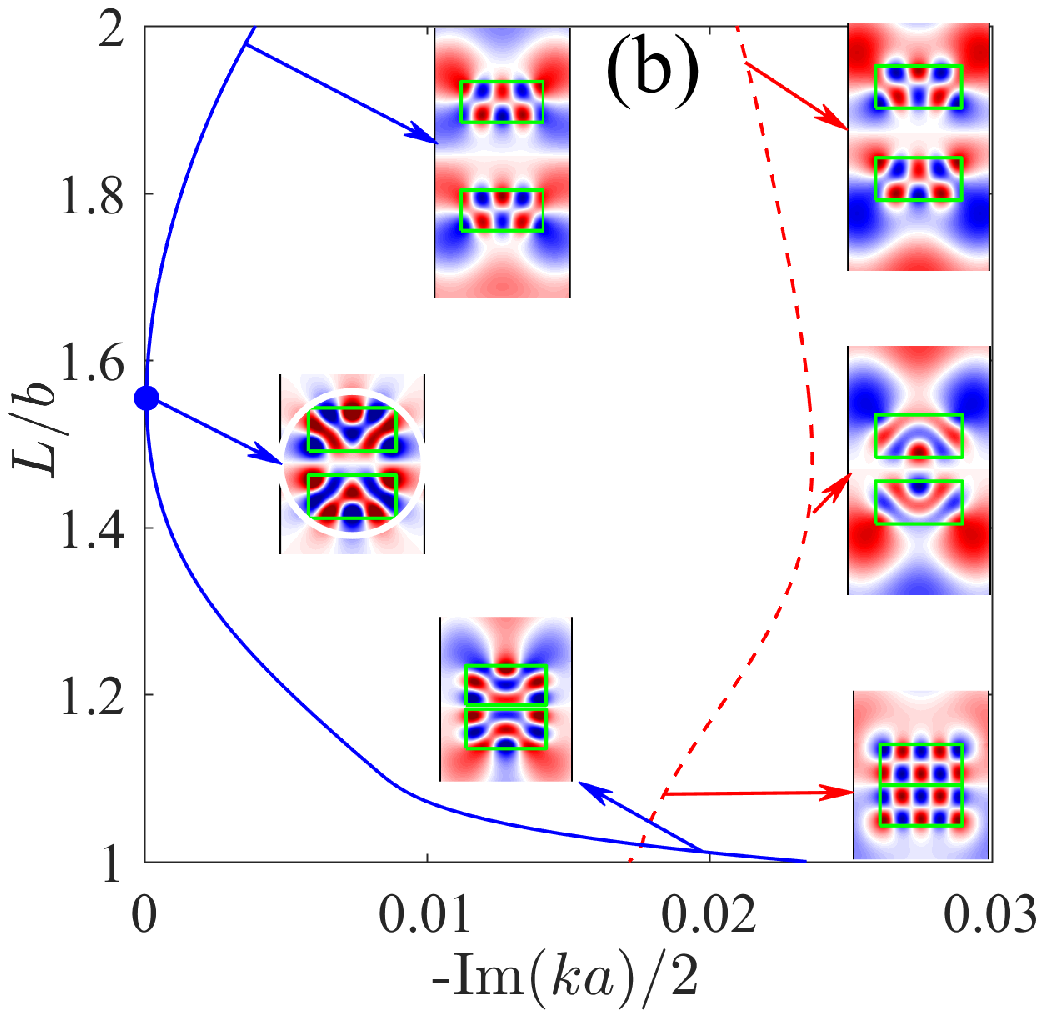}
\includegraphics[width=8cm,clip=]{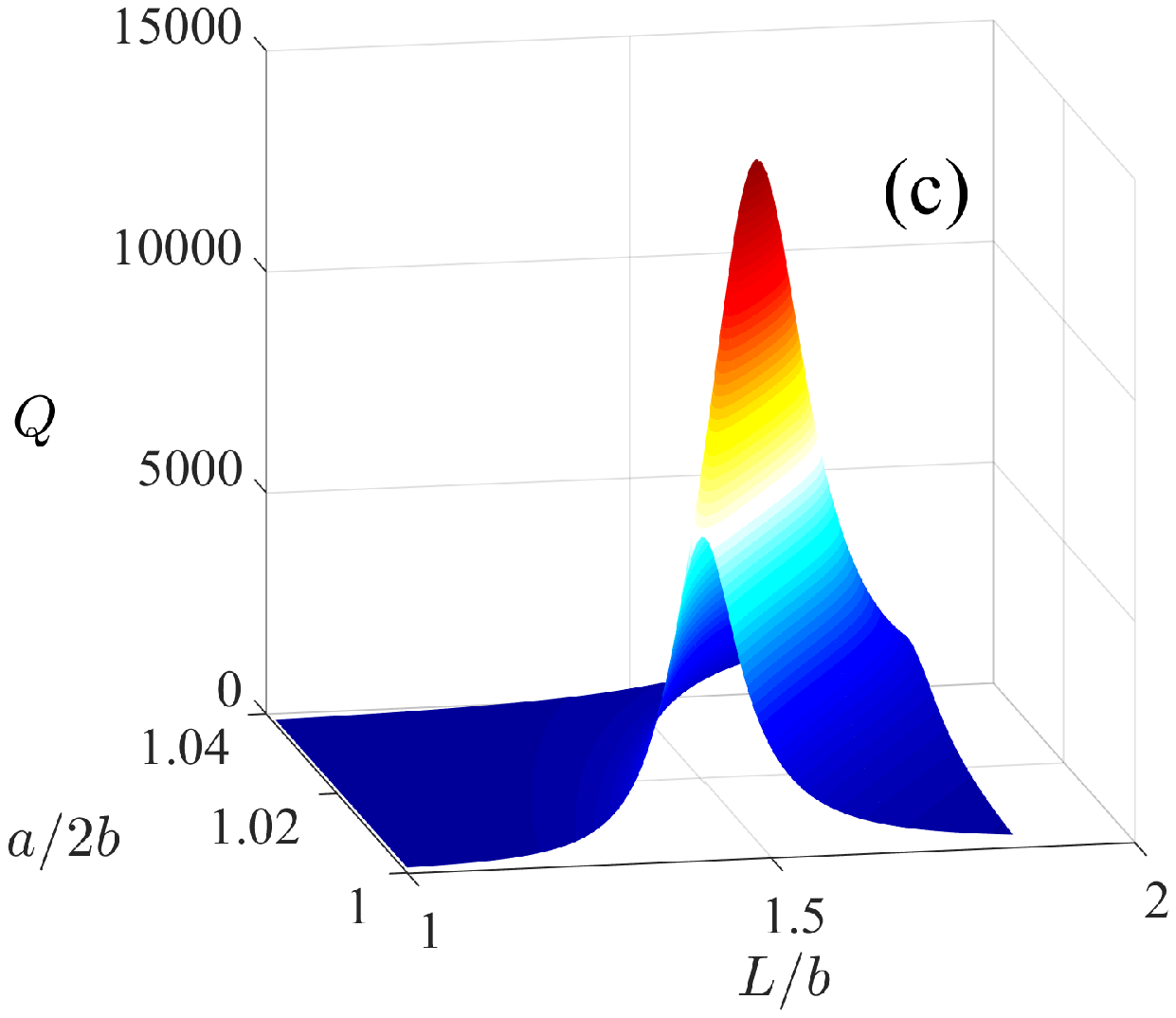}
\caption{(a) Two decoupled TM resonances of opposite symmetry
which are crossing for variation of aspect ratio of isolated wire.
(b) Evolution of imaginary parts of anti-bonding resonances in
traversing with the distance between the wires for $a/2b=1.0152$.
(c) The $Q$ factor vs the distance between the wires and their
aspect ratio. $\lambda=1.55\mu m$ and $a=0.592\mu m$.}
\label{fig3}
\end{figure}
\begin{figure}
\includegraphics[width=4.5cm,clip=]{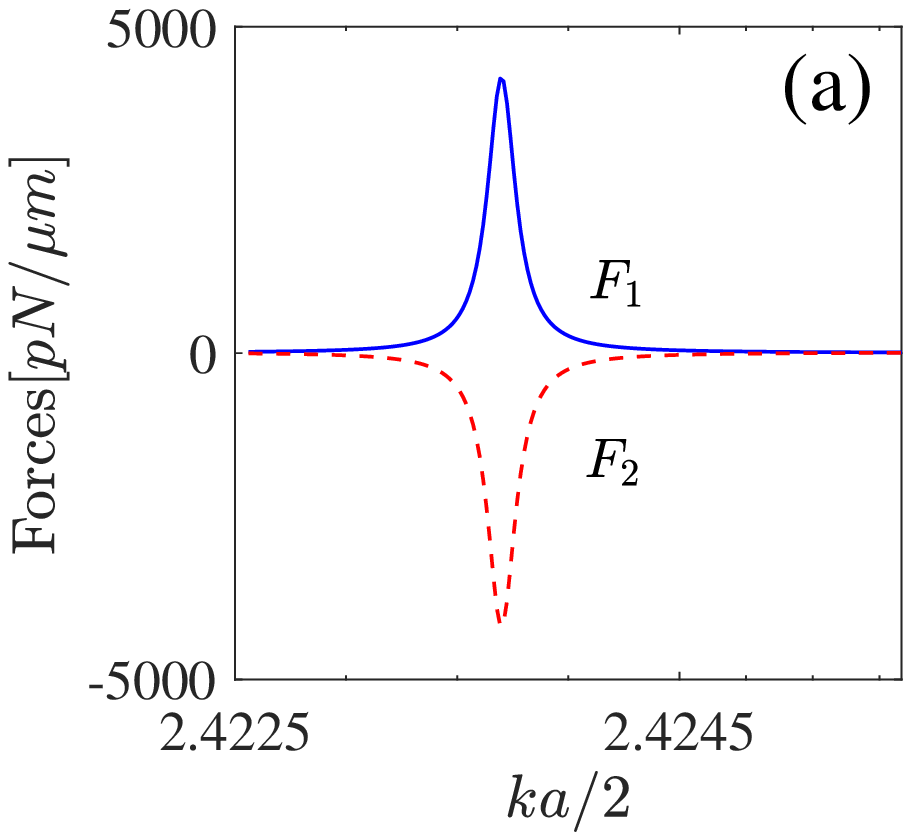}
\includegraphics[width=4.5cm,clip=]{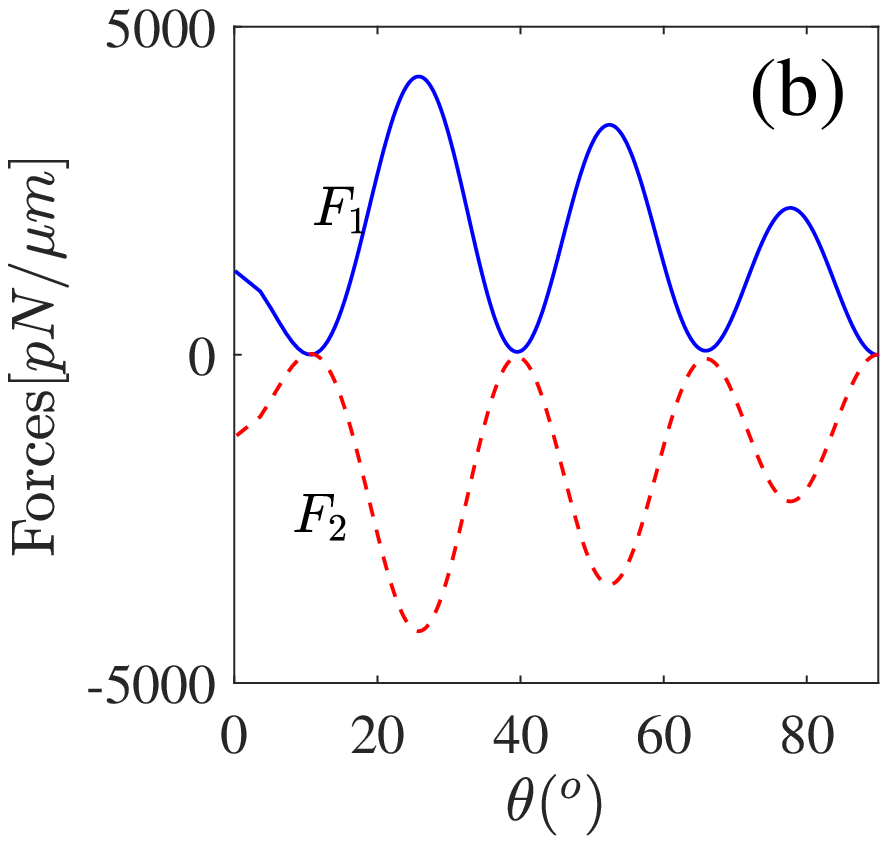}
\caption{(a) The optical forces vs distance between wires at the
vicinity of the anti-bonding  resonance marked in Fig. \ref{fig3}
(b) by closed circle $ka/2=2.423, \theta=25^o$  and $a/2b=1.0152$.
(b) OB force vs the angle of incidence of plane wave $\theta$ at
$L/a=1.5546$ for the power $P_0=1.5mW/\mu m^2$.} \label{fig4}
\end{figure}

In order to achieve unprecedent $Q$ factor the another strategy
was proposed in Ref. \cite{Bulgakov2020a} based on crossing of the
resonances which have opposite symmetry in the isolated wire as
shown in Fig. \ref{fig3} (a). However the presence of the second
wire lifts this symmetry restriction giving rise to a new series
of avoided crossings of resonances. One of such an avoided
crossing is shown in Fig. \ref{fig3} (b) with insets showing
evolution of the anti-bonding resonant modes. We do not show the
bonding resonances because of their undistinguished $Q$ factors.
For variation of the distance $L$ between the wires one of the
anti-bonding resonant modes of two coupled wires reaches
extraordinary small radiation losses. As highlighted in respective
inset in Fig. \ref{fig3} (b) at the point of minimal imaginary
part the anti-bonding resonant mode of two wires with the
cross-section $a=1.19\mu m, b=0.592\mu m, a/2b=1.0152$ and the
distance between wires $L=1.555\mu m$ acquires morphology close to
the morphology of effective isolated wire with cylindrical
cross-section with $m=7$ as highlighted by white circle in inset
of Fig. \ref{fig3} (b). It is remarkable that the $Q$ factor of
this anti-bonding resonance reaches unprecedent value around 15000
at precise tuning of wire's scales: $a/2b=1.0152, L/b=1.5546$ as
shown in Fig. \ref{fig3} (c). Then one can expect giant OB forces
as it was established for two dielectric disks when the frequency
of Bessel beams was resonant to the anti-bonding resonance
\cite{Bulgakov2020}. Indeed Fig. \ref{fig4} show that the OB force
reaches giant values around 4 nano Newtons that exceeds the case
shown in Fig. \ref{fig2} by two orders in value.
\section{Deflection of wires}
Following to Povinelli {\it et al} \cite{Povinelli2005a} we
estimate the deflection of wires $w(z)$ due to optical forces
shown in Fig. \ref{fig4}. The deflection obeys the equation
\begin{equation}\label{defl}
    \frac{d^4w_j}{dz^4}=\frac{12f_j}{ab^3E}, ~~j=1,2
\end{equation}
where $f_j$ are the optical force acting on unit length of the
j-th wire, and $E=169GPa$ is the Young modulus of silicon. In what
follows we disregard optical properties of plates to which the
wires are fixed assuming their refractive index close to air. For
$|w(z)|\ll S$ and applying the boundary conditions $w_j(\pm
S/2)=0$ and $dw_j(\pm S/2)/dz=0$ one can obtain the solution of
Eq. (\ref{defl}
\begin{equation}\label{deflsol}
    w_j(z)=\frac{f_j}{ab^3E}\left[\frac{1}{2}z^4-\frac{S^2}{4}z^2+\frac{S^4}{32}\right].
\end{equation}
At the specific values $a=1.19\mu m, b=0.592\mu m$ and $f_j\approx
\pm 4200 pN$ for the EM power $1.5mW/\mu m^2$ we find that maximal
deflection at the center between plates $w(0)=3.1\times
10^{-8}S^4$ where the length of wires $S$ and deflection $w$ is
measured in microns. Therefore for $S=50\mu m$ we obtain the
maximal deflection is around $150nm$ while the EM power of $1W/\mu
m^2$ propagating along the wires result in  deflection $20nm$
\cite{Povinelli2005a} that considerably yields the case of the
resonant plane wave illumination with the power by three orders
less.

\begin{figure}
\includegraphics*[width=6cm,clip=]{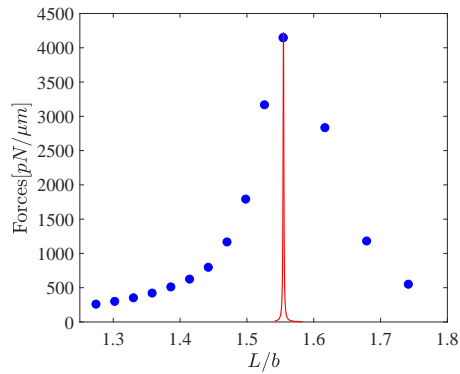}
\caption{The optical forces vs distance between wires at the
vicinity of the anti-bonding  resonance marked in Fig. \ref{fig3}.
Closed circles show the case when the frequency of incident wave
follows to the anti-bonding resonance shown by solid line in Fig.
\ref{fig3} (b). Solid line shows the case when the frequency is
fixed  $ka/2=2.423, \theta=25^o, a/2b=1.0152, L/b=1.555$ that
corresponds to maximal $Q$ factor.} \label{fig5}
\end{figure}

However this estimation of deflection of wires is to be considered
as upper limit. As soon as the wires start to bend the
anti-bonding resonance with extremely high $Q$ factor established
for two parallel wires will go away from extremal point shown in
Fig. \ref{fig3} (b) by, at least, two reasons. The first reason is
related to deviation from optimal distance between the wires. As a
result the optical forces will be strongly decreased as shown in
Fig. \ref{fig5}.
The second reason which also affects the optical forces is related
to the bending of wires which put the problem into the 3d  one.
\section{Funding}
The work was supported by Russian Foundation for Basic Research
projects No. 19-02-00055.
\section{Disclosures}
The authors declare no conflicts of interest.
\bibliographystyle{unsrt}
\bibliography{sadreev}
\end{document}